# Does the Danube exist? Versions of reality given by various regional climate models and climatological datasets

Valerio Lucarini[*] (1,2,3),  Robert Danihlik (2)  Ida Kriegerova (2), Antonio Speranza (1,2)

1. Department of Mathematics and Computer Sciences, University of Camerino, Camerino, Italy
2. CINFAI Unit, University of Camerino, Camerino Italy
3. Department of Physics, University of Bologna, Bologna, Italy

**Abstract**

We present an intercomparison and verification analysis of several regional climate models (RCMs) nested into the same run of the same Atmospheric Global Circulation Model (AGCM) regarding their representation of the statistical properties of the hydrological balance of the Danube river basin for 1961-1990. We also consider the datasets produced by the driving AGCM, from the ECMWF and NCEP-NCAR reanalyses. The hydrological balance is computed by integrating the precipitation and evaporation fields over the area of interest. Large discrepancies exist among RCMs for the monthly climatology as well as for the mean and variability of the annual balances, and only few datasets are consistent with the observed discharge values of the Danube at its Delta, even if the driving AGCM provides itself an excellent estimate. Since the considered approach relies on the mass conservation principle and bypasses the details of the air-land interface modeling, we propose that the atmospheric components of RCMs still face difficulties in representing the water balance even on a relatively large scale. Their reliability on smaller river basins may be even more problematic. Moreover, since for some models the hydrological balance estimates obtained with the runoff fields do not agree with those obtained via precipitation and evaporation, some deficiencies of the land models are also apparent. NCEP-NCAR and ERA-40 reanalyses result to be largely inadequate for representing the hydrology of the Danube river basin, both for the reconstruction of the long-term averages and of the seasonal cycle, and cannot in any sense be used as verification. We suggest that these results should be carefully considered in the perspective of auditing climate models and assessing their ability to simulate future climate changes.

---

[*] Corresponding author, email: lucarini@alum.mit.edu.

# 1. Introduction

Since the climate system features a nonlinear interplay of subdomains, each characterized by complex physical properties (Lucarini, 2002), the definition of strategies for the improvement of numerical climate models is a critical issue in the climate scientists community. This has been recently evidenced by the Project for Climate Model Diagnostics and Intercomparison (PCMDI: http://www-pcmdi.llnl.gov), which has gathered the output of global climate models produced worldwide into a single server and solicited the provision of simple scalar metrics of model performances (Lucarini et al., 2006a).

The auditing of a set of climate models consists of two related, albeit distinct procedures. The first procedure is the intercomparison, which aims at assessing the consistency of the models in the simulation of certain physical phenomena over a certain time frame. The second procedure is the verification, whose goal is to compare the diagnostics of the models to some corresponding observed (or quasi-observed) quantity.

The principle behind regional climate modeling is that, given a detailed representations of natural processes and high spatial resolution that resolves complex topography, land-sea contrast, and land use, a limited area regional climate model (RCM) can generate detailed regional climate information. This information is in principle consistent with the boundary conditions provided by the large scale circulation patterns as described by either global reanalyses data or by general circulation model (GCM) (Wang et al., 2004), within which the RCM is said to be *nested* (Chen and Miyakoda, 1974). Therefore, a RCM is supposed to act as a magnifying lens of the driving model, thus allowing for the representation of small scale features that could not be represented with a coarser resolution (Déqué, 2000). RCMs are based on physical (and possibly, chemical) laws represented by mathematical equations that are solved using three-dimensional grids. They usually include a description of the most important processes affecting the atmosphere and land surface components of the climate system. Many of the processes acting in nature take place on much smaller spatial scales than the resolution of the model grid and cannot be modeled and resolved explicitly. These processes include radiation, convection, surface-atmosphere mass and energy fluxes, turbulent diffusion. Their effects are taken into account using parameterizations, by which the process is represented by deterministic or stochastic relationships between the area or time averaged effect of such sub-grid scale processes and the resolved scale flow.

Regional climate modeling, in addition to the common issues and flaws associated also to global climate modeling and related to the discretisation and parameterization procedures, faces the serious mathematical complication of being a representation of a problem with time-varying boundary conditions. The driving model tends to *enslave* the RCM on time scales depending on the size of the limited domain and constraints at all times the global balances evaluated over the whole domain of the RCM. This implies that future climate projections performed with RCMs may critically depend on the driving global model. Other non-trivial issues arise from the delicate process of matching the boundary conditions at the interface between the coarse and fine resolution models, where rather different spatial and time grids have to be brought to a common ground. Note that today, some models, such as ARPEGE, circumvent some of these problems by allowing for a smooth transition between low resolution-global to high resolution-regional by including a non-uniform grid, whose resolution gradually increases the closer we get to the area of main interest (Courtier and Geleyn, 1988). For more detailed information on RCMs, see Giorgi et al. (2001).

The assessment of the reliability of the current RCMs in the representation of the statistical properties of the hydrological balance of river basins is crucial, because of the relevance of water as a resource and as a source of risks at social, economical, and environmental level (Becker and Grunewald, 2003). Because of the process of latent heat release, biases in the representation of the hydrological balance may in turn strongly effect mesoscale as well as synoptic scale meteorological processes: water is also an active, in a dynamical sense, component of the climate system.

In this study, 12 RCMs and the driving AGCM participating to the PRUDENCE project, and for reference the ERA-40 (ECMWF) and NCEP-NCAR reanalyses, are audited in their representation of the hydrological balance of the Danube river basin. Apart from its primary relevance for the European history, economics, politics, demographics, cultural and environmental heritage, the Danube basin is very interesting from a climatic point of view because it is well within continental Europe while bearing at least a twofold direct relevance to the Mediterranean region. Firstly, the Danube runoff gives a relevant contribution of freshwater flux into the Mediterranean sea (on the average, more than twice the Nile's contribution). Secondly, the Danube depends mostly on precipitated water of Mediterranean origin, because of the geographical position (downwind of the dominant westerlies) and the complex orography of the basin, (Speranza, 2002). When considering the very intense precipitative and disastrous floodings events in central Europe inside and near the Danube basin (Becker and Grunewald, 2003; Stohl and James, 2004), it is well recognized the relevance of the Alps and of the Mediterranean waters in modifying and enhancing the storms of Atlantic origin (Speranza et al, 1985; Tibaldi et al., 1990).

The size is a critical parameter in the choice of a basin as the object of an auditing study of RCMs in terms of water balance. If the basin is so small that only few grid points are contained, we may expect to face noisy data, since the real physical resolution of a model is not given by the distance of neighboring grid-points. On the other hand, if the size of the basin is so large that it represents a relevant portion of the whole domain of the RCMs, we may expect that all diagnostics are heavily constrained by the behavior of the driving model. In this sense, the Danube basin is neither too small, nor too large.

The auditing process of the RCMs can in principle underline characteristic features or flaws, on one side, of the large scale water vapor transport, and, on the other side, of the model representation of some severely parameterized processes occurring in the atmosphere, (e.g. hydrometeors formation and precipitation), at the surface-atmosphere interface, (e.g. evaporation), and inside the soil (e.g. water transport). In nature, the 2D+1D (space and time) fields of main interest for evaluating the hydrological balance are characterized by complex statistical properties, since precipitation features temporal intermittency and spatial multifractal nature (Deidda, 1999; 2000), evaporation and runoff depend very delicately on the local conditions. Therefore, only robust, suitably coarse-grain averaged quantities should be considered in the auditing procedure both in terms of intercomparison and verification, since comparisons at *face value* of model vs. model and, especially, of model vs. observation can be misleading (Accadia et al., 2003). Nevertheless, the procedure of verification, as opposed to the model intercomparison, faces some serious problems if we consider actual observations of precipitation, evaporation, and runoff. Spatial averages of the water balance for the area of interest cannot be determined with reliability from the scattered time series measurements of the surface characteristics. Re-mappings into grids of the climatology of precipitations, allowing for the computation of integrated values, have been proposed only for precipitation (Mitchell and Jones, 2005, Efthymiadis et al., 2006). Consequently, the verification of the climatology can be more naturally performed using a variable of more strict hydrological interest and of much easier experimental access, such as the discharge of the Danube at its Delta. This is possible because, when averages over sufficiently long time scales are considered, the conservation of the water mass implies that the basin integrated value of the difference between precipitation and evaporation must be equal to the basin integrated value of the surface and subsurface runoff, and both must be equal to the river discharge at the end of its course. This quantity is also equal to the convergence of water vapor in the atmosphere over the basin, which is strongly constrained by large-scale meteorological processes. This emphasizes the need for framing the hydrological cycle in meteorological rather than in purely geographical terms.

Several studies focusing on the intercomparison and verification of the hydro-meteorological characteristics and hydrological cycle over specific basins and on various different times scale using GCMs, RCMs, and reanalyses have been recently proposed. We may mention, certain of making an incomplete list, the papers by Roads et al. (1994), Lau et al. (1996), Gutowski et al. (1997), Betts et al. (1999, 2003), Roads and Betts (1999), Hagemann et al. (2004), Hirschi et al. (2006). In these analyses, a combination of techniques such as the evaluation of integrated values of the convergence of atmospheric water vapor, of precipitation minus evaporation, of runoff, of the variation of terrestrial water have been employed for large scale basins. In particular, the remarkable and encyclopedic paper by Hirschi et al. (2006) provides a rather complete analysis of the mean climatology of the water balance by diagnosing 37 (!) world major river basins, including the Danube, and providing a dataset of basin-scale water-balance, which is available for download through Internet.

We believe that this paper, which presents some of the results of the 2006-2007 EU INTERREG IIIB-CADSES project HYDROCARE (http://www.hydrocare-cadses.net), can give a novel contribution by presenting a more detailed analysis of the specific performances of a rather large number of models in their representation of the statistical properties of the water balance data over the Danube basin. Apart from the usual evaluation of the seasonal cycle of the water balance, for all datasets we compute in two independent ways its long-term averages and the 95% confidence interval of the annual average; we analyze statistically the precipitation-evaporation feedback; we evaluate the impact of North Atlantic Oscillation (NAO) and El-Niño-Southern Oscillation (ENSO) variability, and we evaluate quantitatively the statistical relationship between the water balance of all RCMs and what predicted by the AGCM they are driven by, in order to estimate the improvement of the information obtained by means of the nesting procedure.

The paper is structured as follows. In Section 2, basic information about the Danube Basin, the data considered in this study and the concepts behind the diagnostics tools employed in the auditing are presented. In Section 3 and 4 we present and discuss the main results on the intercomparison and verification of the models, respectively regarding the yearly and the monthly climatology of the precipitation, evaporation, water balance and runoff. In Section 5 we draw our conclusions and present perspective for future research in this field.

## 2. Data and methods

**2.a Notes on the Danube river basin**

The location of the Danube river basin is shown in Figure 1. The Danube river is about 2850 km long, its 1961-1990 average discharge is about 6600 $m^3.s^{-1}$, and its basin covers about 800,000 $km^2$ (Global Runoff Data Center, Germany, http://grdc.bafg.de/) in 18 countries (Germany, Austria, Slovakia, Hungary, Croatia, Serbia & Montenegro, Romania, Bulgaria, Moldova; Ukraine, Poland,

Czech Republic, Switzerland, Italy, Slovenia, Bosnia-Herzegovina, Albania, FYR of Macedonia) – about one-third of continental Europe outside Russia. The Danube is the most international river basin in the world and is the subject of a huge number of international projects, including the UNDP/GEF Danube Regional Project (http://www.undp-drp.org/drp/index.html) and International Commissions, including the International Commission for the Protection of the Danube River (http://www.icpdr.org/), just to name a few. The Danube catchment encompasses continental climate, as it is land-dominated by advection from the surrounding land areas (especially central and eastern regions). Only the western parts of the upper basin, in Germany, are influenced by the Atlantic climate and the south-west of the basin (ex-Yugoslavian countries) by the Mediterranean climate. The Alps in the west, the Dinaric-Balkan mountain chains in the south and the Carpathian mountain bow in the eastern center are distinctive morphological and climatic regions and barriers. These mountain chains receive the highest annual precipitation (1,000-3,200 mm per year) while the inner and outer basins (Vienna basin, Pannonian basin, Romanian and Prut low plains), the lowlands and the delta region are very dry (350-600 mm per year).

**2.b Datasets**

The following data sources relative to the 1961-1990 time-frame have been used for the purposes of our analysis:

1. Daily values of Runoff (R), Precipitation (P), Evaporation (E), from 9 Regional Climate Models (RCMs) (see Table1), driven at the boundaries by the HadAM3H AGCM in the A2 scenario (see below). Two RCMs (DMI and SMI) provide output data resulting from runs where various resolutions are adopted for the models, so that a total of 12 runs are considered. The grids of all models are defined in terms of a rotated coordinate systems, allowing for having quasi-square grids over Europe in both the natural and the angular metrics. The whole of Europe is well-inside the domain of the RCMs, as shown for a typical case (PROMES UCM), in Fig. 1c. The data have been produced in the context of the 5th Framework Programme of the EU project named PRUDENCE and have been obtained through the PRUDENCE website, which contains information about the RCMs (http://prudence.dmi.dk).
2. Daily values of precipitation (P) and evaporation (E), from the HadAM3H atmospheric general circulation model (AGCM), forced (scenario A2) by the observed and reconstructed boundary conditions at the atmospheric interfaces (see Table 2). HadAM3H is an improved version of the atmospheric component of the Hadley Centre coupled atmospheric-ocean global circulation model HadCM3 (Gordon et al. 2000). The data have been obtained directly from MetOffice, Hadley Centre, UK.
3. Daily values of runoff (R), precipitation (P), Evaporation (E) from 2 major reanalysis datasets (see Table 2): the ERA-40 reanalysis released by the European Center for Mid-Range Weather Forecast (ECMWF) (Simmons and Gibson 2000) and the NCEP-NCAR reanalysis produced by the National Center for Environmental Prediction (NCEP), in collaboration with the National Center for Atmospheric Research (NCAR) (Kistler et al. 2001). For the latter reanalysis E data has been obtained straightforwardly from the Latent Heat Flux data. The data have been downloaded from http://data.ecmwf.int/data/d/era40_daily/ (ERA-40) and http://www.cdc.noaa.gov/cdc/reanalysis/reanalysis.shtml (NCEP-NCAR)
4. Daily discharge (D) of the Danube river at the near-sea Ceatal Izmail station (45.22°N, 28.73°E) (see Table 2). Data have been obtained from the Global Runoff Data Center (GRDC), Germany (http://grdc.bafg.de/).

Note that the PRUDENCE datasets comprises outputs of other RCMs; for matter of consistency and with the purpose of limiting spurious effects due to the boundary conditions, we have selected the largest subset of runs that are nested in the same simulation of a single AGCM.

**2.c Notes on the Theoretical framework**

We proceed along the lines of Peixoto and Oort (1992). By imposing mass conservation for water, we obtain the following expression for the local balance equation of the terrestrial water:

$$\frac{\partial S}{\partial t} = P - E - R \qquad (1)$$

where S represents the terrestrial water storage per unit area, *R* the runoff (including the surface and the subsurface runoff of the area), *P* is precipitation, *E* is evaporation. In this work by evaporation we mean the total evaporation, thus including transpiration. When considering an atmospheric column stretching from surface to the top of the atmosphere, we have that the balance equation can be written as follows:

$$\frac{\partial W}{\partial t} = -\vec{\nabla}_H \cdot \vec{Q} - P + E \qquad (2)$$

where $W$ represents the column storage of water vapor, $\vec{Q}$ is vertically integrated two-dimensional water vapor flux, and $\vec{\nabla}_H \cdot$ is the horizontal divergence operator.

On time scales $T$ which are long compared to the average residence time of water in the atmosphere (~ 10 days), to the duration of the temporary storage of water in form of snow cover (~ few months at most), and to the seasonal duration (3 months), say $T \geq 1y$, we can safely assume that:

$$\left\langle \frac{\partial S}{\partial t} \right\rangle_T \equiv \frac{1}{T} \int_t^{t+T} \frac{\partial S}{\partial t} dt \approx 0 \qquad (3a)$$

$$\left\langle \frac{\partial W}{\partial t} \right\rangle_T \equiv \frac{1}{T} \int_t^{t+T} \frac{\partial W}{\partial t} dt \approx 0 \qquad (3b)$$

where the square brackets represent the operation of time averaging. It follows that:

$$\langle P \rangle_T - \langle E \rangle_T \approx \langle R \rangle_T \qquad (4a)$$

$$-\langle \vec{\nabla}_H \cdot \vec{Q} \rangle_T \approx \langle R \rangle_T \qquad (4b)$$

This implies that, when considering the long-term average, the difference between precipitation and evaporation equals the surface and subsurface runoff, which also equals the convergence of the atmospheric water flux. If we integrate spatially Eqs. (4a-b) over the entire geographical region $A$ corresponding to the hydrological basin of a river and impose the conservation of water, we obtain the following basic form of hydrological balance:

$$\int_A d\sigma \left( \langle P \rangle_T - \langle E \rangle_T \right) = \int_A d\sigma \langle B \rangle_T \approx -\int_A d\sigma \langle \vec{\nabla}_H \cdot \vec{Q} \rangle_T \approx \int_A d\sigma \langle R \rangle_T \approx \langle D \rangle_T \qquad (5)$$

where $B = P - E$ is the net balance and $D$ is the actual river discharge into the sea. Note that, considering using Gauss' integral theorem, we also have that the time-averaged river discharge equals the time average of the net incoming atmospheric flux of water through the vertical boundaries of the atmospheric region bounded below by the region $A$: in a way, the river flows down from the sky. Equations (5) form the basis of the diagnostic study presented in this paper.

We now describe the data processing procedure adopted in this work for each model output allowing for the independent estimates of the daily time-series of the basin-integrated values of the precipitation, evaporation, and runoff fields, to be used for studying the validity of Equation (5) in the analyzed datasets. MATLAB 7.0.4®, MS Office Excel® and ArcGIS 9.0® software packages have been used with customized routines. The first step after data collection grounds in reading the converting the input data (*P*, *E*, *R*), from the .NC format to .MAT format, which is the standard format for MATLAB 7.0.4® software. This is accomplished by using the netCDF tool available at http://www.marine.csiro.au/sw/matlab-netcdf.html. In the MATLAB 7.0.4® environment, the numerical manipulations of the raw data are performed in order to obtain the climatology of the gridded data. Then, the outputs are transferred to MS Office Excel®, where DBF files are created, and, subsequently, these are imported into ArcGIS 9.0® software. In GIS environment, a new point layer of all characteristics is created (Figure 1a).

The final step – the numerical integration of the fields - can be performed by adopting two conceptually distinct strategies. The first approach relies on considering, for a given grid-point, the analyzed field as constant within the corresponding grid-cell centered on the grid point. Then, all the contributions coming from cells contained inside the geographical domain of the river basin are summed up with a weight corresponding to the area of the cell, while the contributions from the boundary cell (not entirely contained in the basin) are weighted with the portion of the area of the cell inside the basin. The cells can be defined in general terms by producing the Voronoi or Thiessen tessellation (Okabe et al., 2000) of the grid, where the grid-cell corresponding to a grid-point is defined as the set of points that are closer to that grid-point than to any of the other ones. In the cases considered in this study, since the grids are locally quasi-rectangular (actually, quasi-square) in the natural metric, the Voronoi tessellation is such that the grid cell corresponding to the grid-point $(j,k)$ is basically a rectangle with corners given by the combinations of the grid points $[(j\pm 1, k\pm 1)+(j,k)]/2$. This approach guarantees, apart from second order numerical approximations, that the total water flux is computed exactly (Figure 1b). The second approach, which is computationally more expensive, relies on interpolating (with various algorithms, such as splines) the gridded data up to the same resolution of the basin perimeter and then following the above mentioned strategy, where now by definition no boundary cells are presents (cells are either inside or outside the basin). Note that this second approach aims at producing a detailed (but somewhat arbitrary) representation of the 2D field, but does not ensure the a-priori consistency of the integrated flux. We have tested that the numerical space integrations performed by adopting various versions of the two above-mentioned strategies give results that are in satisfactory agreement (within 1-2%), at least in the sense that these corrections are negligible with respect to the later described inter-model discrepancies. The results presented in this paper have been obtained with the Voronoi tessellation approach.

**2.d From observed discharge back to runoff.**

Based on the methodology of Hagemann et al. (2004), we can perform an approximate, empirical reconstruction of monthly *quasi-observed* runoff values starting from observed monthly river discharge $D$. Starting from the daily fields, we can define the monthly-averaged time series of the basin integrated $P$, $E$, $B$, and $R$ fields as follows:

$$\overline{\Phi}_k^j \equiv \int_A d\sigma \langle \Phi \rangle_{Y=k, M=j}, \qquad (6)$$

where $\Phi$ is any of the field $P$, $E$, $B$, or $R$ and, with obvious meaning of symbols, the time averaging on the right hand side of Equation (6) is performed over the $j^{th}$ calendar months of the $k^{th}$ of 30 years. Therefore, $\overline{\Phi}_k^j$ is a time series composed of $30 \times 12 = 360$ elements. For a certain catchment, Hagemann et al. (2004) assume that the relationship between basin integrated monthly mean runoff $\overline{R}_k^j$ and discharge $D_k^j$ can be approximated as follows:

$$D_k^j = a_j \overline{R}_k^{j-L}, \qquad (7)$$

where L is an average lag between $\overline{R}_k^j$ and $D_k^j$, and the periodic factors $a_j$ take care of describing the area integrated *transfer function* from the basin to the Delta of the river. Hagemann et al. (2004) obtains the optimum values of L and $a_j$ for each of the 12 calendar months from the 15 year time series (1979-1993) of simulated monthly total runoff and discharge values using a least square fit, allowing for integer lag values only. The input are obtained with the HIRHAM4 RCM (Hagemann and Dümenil Gates., 2001), forced by ERA-40 boundary conditions, feeding runoff data into a hydrological discharge model (Hagemann and Dümenil, 1999), so that the fit is nothing but the approximate solution of an inverse problem. For the Danube catchment, the optimum lag is found to be *L=1* (month), while the values of $a_j$ are reported in Table 3. Thus, Equation (7) becomes

$$D_k^j = a_j \overline{R}_k^{j-1} \qquad (8)$$

for month *j* (cyclic). Obviously, the output of this optimization procedure depends strongly on the hydrological discharge model adopted and (much more weakly) on the time duration and input data. Anyway, in the perfect hydrological model scenario (which is false, of course) and neglecting the other effects, Equation (7) allows us to obtain from a time-series of observed discharges the corresponding hypothetical (in our terms, reconstructed) basin integrated runoff time-series. We then adopt this simplifying working hypothesis and generate such quasi-observed monthly time-series for 1961-1990, taking care of making a minor adjustment: we divide the $a_j$ presented in Table 3 by the same factor f=1.06, so that the 1961-1990 long term averages of $D_k^j$ and $\overline{R}_k^j$ coincide, thus respecting automatically Equation (5) for $T=T_{MAX}=30$ years.

## 3. Results: Yearly Hydrological balance

The considered datasets are analyzed in terms of their representation of the hydrological balance over the Danube basin by focusing on the long-term mean, on the interannual variability, and on the intra-annual variability (e.g., the seasonal cycle).

For each model, we define, by suitably averaging over the calendar years, the yearly time series of the accumulated basin integrated fields as follows:

$$\overline{\Phi}_i \equiv 1y \times \int_A d\sigma \langle \Phi \rangle_{i+1960}, \qquad (9)$$

where $\Phi$ is any of the field *P, E, B,* or *R* and the time averaging on the right hand side of Equation (8) is performed over the calendar year indicated as the lower index, and $1y$ is the time-length of 1 year. Therefore, $\overline{\Phi}_i$ is a time series composed of 30 elements. Using standard statistics, we have that the best estimate of the long-term mean of $\overline{\Phi}_i$ can be written as:

$$\hat{\mu}(\overline{\Phi}_i) = \frac{1}{30} \sum_{i=1}^{30} \overline{\Phi}_i = \frac{1}{30} \times 30y \times \int_A d\sigma \langle \Phi \rangle_{T_{MAX}}, \qquad (10)$$

where the second equality ($T_{MAX}$ is whole 1961-1990 time frame) is rather obvious, while the best estimate for the standard deviation of $\overline{\Phi}_i$ can be written as:

$$\hat{\sigma}(\overline{\Phi}_i) = \left[\frac{1}{29}\sum_{i=1}^{30}(\overline{\Phi}_i - \hat{\mu}(\overline{\Phi}_i))^2\right]^{1/2}, \quad (11)$$

Assessing the mutual consistency between the climatologies provided by the various models entails comparing their estimates of $\hat{\mu}(\overline{\Phi}_i)$, $\hat{\sigma}(\overline{\Phi}_i)$, *and* considering the statistical uncertainties associated with such estimates.

For all models and for all choices of the fields $\Phi$, the time series $\overline{\Phi}_i$ are compatible with the null hypotheses of white noise. In all cases, the values of the estimates of the lagged correlations are smaller than 0.35 for all time lags $\geq 1$ year, whereas the corresponding 95% confidence interval for a synthetic white noise time series of the same length is about [-0.4, 0.4]. Similarly, the yearly time series do not show any statistically significant correlation with the North Atlantic Oscillation index of the HadAM3 A2 driving model (Bojariu and Giorgi, 2005), nor with the various El Niño-Southern Oscillation indexes (see http://www.cpc.ncep.noaa.gov/data/indices/), which, being related to oceanic surface temperatures, are basically an input to the driving model. Therefore, for each model the confidence interval for the estimate of the climatological mean is centered on the quantity $\hat{\mu}(\overline{\Phi}_i)$ (the estimator can be assumed to be unbiased) and its half-width can be approximated as $2\hat{\sigma}(\mu(\overline{\Phi}_i)) \approx 2\hat{\sigma}(\overline{\Phi}_i)/\sqrt{29} \approx 0.37\hat{\sigma}(\overline{\Phi}_i)$, where $\hat{\sigma}(\mu(\overline{\Phi}_i))$ is the standard deviation of the mean. Actually, other approaches, such as block-bootstrap methods (Wilks, 1997; Lucarini et al., 2006b), give estimates of $\hat{\sigma}(\mu(\overline{\Phi}_i))$ that are consistent within 20%. If the confidence intervals of two models do not overlap, we can say that their climatological means are not statistically consistent. If we had longer runs, we could restrict progressively the confidence interval of the climatological mean. Whereas the length of the considered simulations allows for a reliable statistical interpretation of the model discrepancies in the description of intra-seasonal variability, the same is not true for the interannual variability. However, by treating the variance of $\hat{\sigma}(\overline{\Phi}_i)$ as a $\chi^2$-distributed random variable with 29 degrees of freedom and by choosing a confidence level of 95%, we obtain in all cases a confidence interval spanning around 30-40% of the variance itself. Therefore, the length of the time series is not enough for a detailed statistical assessment of the models' discrepancies in the description of interannual variability.

**3.a Balance, Precipitation, Evaporation**

We start by considering $\Phi = B = P - E$. Results are presented in Figure 2. The scatter plot portraits the interannual variability $\hat{\sigma}(\overline{B}_i)$ of the integrated water balance vs. the 95% confidence interval of the best estimate of the yearly average $\hat{\mu}(\overline{B}_i)$, given by $[\hat{\mu}(\overline{B}_i) - 2\hat{\sigma}(\mu(\overline{B}_i)), \hat{\mu}(\overline{B}_i) + 2\hat{\sigma}(\mu(\overline{B}_i))]$.

The graphical result displays that the RCMs are clustered into three quite distinct groups. The statistics of RACMO KNMI, of HIRHAM METNO, and of the two versions of SMHI agree with the statistics of HadAM3, the three versions of DMI, CHRM ETH, REMO and CLM GKSS greatly underestimate, by factors up to 50%, both $\hat{\sigma}(\overline{B}_i)$, and, especially $\hat{\mu}(\overline{B}_i)$. Finally, two models (ICTP and PROMES UCM) overestimate these quantities up to 30%. When considering the 95% confidence interval of $\hat{\mu}(\overline{B}_i)$, we have that the second and third group of RCMs are not consistent with the driving model by many standard deviation, so that their climatologies of yearly water balance are definitely not consistent. Since all the RCMs are driven by the same run of HadAM3, such discrepancies seem rather peculiar. We also note that increases in the resolution do not significantly alter the performance of the DMI and SMHI models.

If we consider Equation (5) and plug in $T=T_{MAX}$ and recall Equation (9), we have that the long-term average of the Danube discharge should provide a verification value for $\hat{\mu}(\overline{B}_i)$ and is reported in Figure 2. From the GRDC dataset, we have that $\langle D \rangle_{T_{MAX}} \approx 6600 m^3 s^{-1}$. We observe that the water balance of the driving model is in a remarkable agreement with this verification value, with the 95% confidence level of $\hat{\mu}(\overline{B}_i)$ well including the value of $\langle D \rangle_{T_{MAX}}$. The RCMs whose statistics is clustered around HadAM3's also agree well with the verification value, with the model HIRHAM METNO giving an almost exact value. Therefore, the statistics of the two other groups of RCMs have large *wet* or *dry* biases.

When considering the two reanalyses, we have that neither agrees with the observational data. In particular, the 95% confidence level of $\hat{\mu}(\overline{B}_i)$ for the ERA40 reanalysis intersects zero, with actually several years of the time series $\overline{B}_i$ well below zero. A negative value for $\hat{\mu}(\overline{B}_i)$, or for even one of the entries of $\overline{B}_i$ is clearly absurd, since it would imply that the Danube basin is a net exporter of water, so that the presence of a river would be impossible. Most sea basins are net exporters of water , while river basins, by definition, have to be net importer of water (Peixoto and Oort, 1992). This suggests that the ERA40 datasets faces some serious problems in terms of water budget, at least in this area of the world.

Further information on the representation of the hydrological cycle can be obtained by analyzing the disaggregated fields of precipitation and of evaporation. Results are presented in

Figure 3. We first observe that the width of the 95% confidence interval of $\hat{\mu}(\overline{P}_i)$ is in all cases much larger, by almost an order of magnitude, than that of $\hat{\mu}(\overline{E}_i)$, thus meaning that the precipitative fields have a much larger interannual variability, and their statistics is *less constrained*. We see a very large span of values, no clustering similar to the previous case is present. Anyway, we see qualitatively a positive correlation among models between $\hat{\mu}(\overline{P}_i)$ and $\hat{\mu}(\overline{E}_i)$, *i.e.*, some models have high values of $\hat{\mu}(\overline{P}_i)$, which tend to be compensated (in terms of net balance) by high values of $\hat{\mu}(\overline{E}_i)$, while other models have consistently a weak hydrological cycle both in the precipitation and evaporation channel. Note that the two models with highest (and virtually indistinguishable) values of $\hat{\mu}(\overline{B}_i)$, the PROMES UCM and the ICTP, are in this graph in total disagreement, with the former featuring the weakest of all hydrological cycles, with a super-low total evaporation, and the latter featuring one of the strongest hydrological cycles, with a very high precipitation. Note also that some dry models, such as the DMI family or the REMO model, feature precipitation statistics which are in agreement with HadAM3, but have much stronger evaporation. The RCMs which are in good agreement with HadAM3 in terms of net balance - and with the observative constraint, here represented with a straight line parallel to the bisectrix – are actually quite sparse in this graph. The clustering observed in Figure 2 is here represented by a sort of banded structure, parallel to the bisectrix, in the position of the model-representative dots. Note that, when considering different version of the same model, we have, in both the cases of SMHI and DMI models, that the increase in the resolution causes an enhanced precipitation (due to small scale features) but at the same time also a compensating enhanced evaporation, so that the net water balance, which is determined by the large scale atmospheric water influx into the domain and on the efficiency of large scale precipitation, does not depend on the resolution, as can be observed in Figure 2. The large span of the whole sets of RCMs and the effect of the resolution as depicted in the precipitation-evaporation plane (Fig. 3) – as opposed to the clustering when water balance is considered (Fig. 2) - may be interpreted as a strong model sensitivity to small scale and heavily parameterized features, such as soil and vegetation properties, localized (*e.g.* topography-induced) and convective rains efficiency, which do not alter the global balance but effect in a relevant fashion the total intensity of the hydrological cycle and are relevant in the separate issue of analyzing flood events.

Looking at the time-correlations of the $\overline{B}_i$, $\overline{P}_i$ and $\overline{E}_i$ time series, we find some hints regarding the observed properties of the mean values. Let's consider Table 4. Considering that the 95% confidence level for the null hypothesis of absence of correlation is about 0.4, all the RCMs have very a high degree of correlation with the driving model for the $\overline{B}_i$, in some cases reaching or exceeding 0.9. Such correlations may be related, in the first place, to the fact that the RCMs follow the variability of the driving model in fuelling the influx of water vapor into the area. At the same time, from Table 5, we observe that RCMs and HadAM3 feature a very strong internal precipitation-evaporation feedback, since the correlation of the $\overline{P}_i$ and $\overline{E}_i$ time series is very high and positive, except for one model (PROMES UCM), where, unexpectedly, such a correlation is statistically not significant. This shows that a positive mechanism is set up when local processes are considered: higher precipitation brings to wetter soil and so to higher evaporation, which increases the water content of the atmosphere. This feedback does not preclude the stabilization of the net hydrological balance, since the precipitation and evaporation anomalies tend to cancel out. Note that, when considering the reanalyses, while the NCEP/NCAR datasets is consistent with that picture, ERA40 shows a negative correlation (at the edge of 95% statistical significance) between the $\overline{P}_i$ and $\overline{E}_i$ time series, thus reinforcing the idea that in this latter reanalysis some serious issues are present in modeling water processes. Nevertheless, as seen in Figure 3, the precipitation variability is much larger and dominates in all RCMs and in HadAM3, when balance is considered (this applies also for the PROMES UCM model). Thanks to the interplay of the driving model's constraint and of the internal mechanism, as we can observe from Table 4, also the RCMs $\overline{P}_i$ and $\overline{E}_i$ time series are very strongly correlated with those of $\overline{P}_i$ of the HadAM3, except for the case of the evaporative data of PROMES UCM, in agreement with what observed above.

Therefore, we may deduce that, while the RCMs actually act as strongly constrained downscaling models, at the same time, once outputs are upscaled via spatial integration procedure on a finite - not too large, not too small domain, as discussed earlier - domain, information may be, and actually in most cases is, degraded.

**3.b Runoff**

Further information on the performance of the models can be obtained using a diagnostic procedure relying on Equation (5), which states the equivalence, when long time-averages are considered, between the water balance obtained by integrating the difference between the precipitation and the evaporation fields and that obtained by integrating the runoff, which results as output of the (severely parameterized) water dynamics within soils. Unfortunately, runoff data for the HadAM3 model have not been available. In Figure 4 we present the scatter plot where the abscissa of each dot representative of a dataset is given by $\hat{\mu}(\overline{R}_i)$, while the ordinate is given by $\hat{\mu}(\overline{B}_i)$. The observative constraints are also shown. Note that in all cases the width of the 95% confidence 1   interval for $\hat{\mu}(\overline{R}_i)$ is about 20% smaller than that of $\hat{\mu}(\vec{B}_i)$, the reason of the smaller variability of
2   $\hat{\mu}(\overline{R}_i)$ being that the runoff is mediated by the soil processes, which introduce an effective low-pass
3   filtering. Equation (5) implies that, for a matter of consistency, the dots should line up along the
4   bisectrix - $\hat{\mu}(\overline{B}_i) = \hat{\mu}(\overline{R}_i)$ - and consistency between the statistics of the yearly averages cannot be
5   rejected when the 95% confidence intervals presented in figure intersect it. If $\hat{\mu}(\overline{B}_i) < \hat{\mu}(\overline{R}_i)$, we
6   have that on the long term the soil component of the RCMs generates water, whereas destruction is
7   implied if the other sign of the inequality holds. For the CLM GKSS model, the two estimates of
8   water balance are definitely not consistent with each other, with $\hat{\mu}(\overline{R}_i)$ larger than $\hat{\mu}(\vec{B}_i)$ by over
9   20%. Other RCMs, such as ICTP and SMHI50, have representative dots slightly off the bisectrix
10   $\hat{\mu}(\overline{B}_i) < \hat{\mu}(\overline{R}_i)$ - but still have statistically significant agreement between the two estimates. When
11   looking at the reanalyses, we have that in both cases the disagreement between $\hat{\mu}(\vec{B}_i)$ and $\hat{\mu}(\overline{R}_i)$ is
12   very large, with a very strong case of $\hat{\mu}(\overline{B}_i) < \hat{\mu}(\overline{R}_i)$. In particular, the ERA40 dataset features
13   $\hat{\mu}(\overline{R}_i) > \langle D \rangle_{T_{MAX}}$, whereas $\hat{\mu}(\overline{B}_i) \approx 0$, so that a river of about the same size as the actual Danube
14   seems to be created. The fact that the ERA40 datasets does not conserve water has already been
15   observed by Hagemann et al. (2005). In all cases, where there is statistical disagreement between
16   $\hat{\mu}(\overline{R}_i)$ and $\hat{\mu}(\overline{B}_i)$, we may guess that serious issues in the representation of the soil processes are
17   present and need attention.

## 4. Results: Seasonal Cycle of the Hydrological balance

22 As a dual analysis to what presented in the previous section, we show some results regarding the
23 intra-annual variability of the variables relevant for the hydrological balance of the Danube basin.
24 In particular, we analyze the *seasonal cycle*: for a discussion on various definitions of the seasonal
25 cycle, see *e.g.* Lucarini et al., (2004). We hereby identify the seasonal cycle with the usual long-
26 term monthly averages of the records, so that, following the definitions contained in Equation (5),
27 the seasonal cycle of the basin-integrated monthly accumulated field $\Phi$ is given by:

$$\overline{\Phi}^j \equiv \frac{1}{30} m^j \times \sum_{k=1}^{30} \overline{\Phi}_k^j , \quad j = 1,...,12 , \quad (12)$$

where the index $j$ refers to the $j^{th}$ month, and $m^j$ is its length. Obviously, when considering the different length of the months, we have that the suitably weighted sum of $\overline{\Phi}^j$ gives $\hat{\mu}(\overline{\Phi}_i)$ for $\Phi = P, E, B, R$.

The seasonal cycle of basin-integrated precipitation is presented in Figure 5a. We first observe that all RCMs, the HadAM3 model, and the ERA40 reanalyses qualitatively agree on the overall features, even if for all months the span is well over 50% of the ensemble mean of RCMs (not in figure). The quantitative agreement for the climatology of the early summer precipitations is especially problematic, probably due to the delicate model-dependent tuning of the convective processes. With this respect, it is notable that the NCEP/NCAR reanalysis gives very high values for the May-to-August precipitations, which are month-wise larger by a factor of about 2 than the second largest entry. The plot contained in Figure 5b describes the occurrences of RCMs having the absolute maximum (blue) or minimum (red) of the seasonal cycle in the corresponding month. Precipitations peak in the early summer months and in November, which gives more often a secondary maximum. This is a typical feature of the observed climatology of the precipitation of the Danubian region, as shown by Brunetti et al. (2006), albeit on a more limited domain. The minima of precipitation are also clustered in two periods of the year, namely February and the summer months. The latter is a signature of the influence of the typical Mediterranean summer.

The seasonal cycle of evaporation is presented for all datasets in Figure 6a. Again, we have that all models qualitatively agree, even if the span of the values is over 50% of the ensemble mean of RCMs (not in figure) for all months. Figure 6b shows a characteristic insolation-dependent pattern, where the maxima of evaporation are reached in all cases around the summer solstice and the minima are realized around the winter solstice. It is notable that the PROMES UCM model, which has been shown in the previous section to be problematic in the representation of the precipitation-evaporation feedback, features the smallest and somewhat delayed seasonal cycle for evaporation. Compared to the other datasets, the NCEP/NCAR reanalyses, instead, features the strongest evaporation in virtually all months.

Combining the information on precipitation and evaporation, we obtain the seasonal cycle of the water balance, which is depicted in Figure 7a. The agreement between the RCMs is better for this diagnostics, as could be guessed by the results of the previous section, since the biases in the precipitation and evaporation fields tend to even out. It is remarkable that – see Figure 7b - all RCMs have the largest positive water balance in November and the largest negative balance in July, which again shows that the water balance is structurally a more robust variable. Figure 7a clearly shows a major problem for the NCEP/NCAR reanalyses, which is in total disagreement with the other datasets. The water balance is positive and peaks in the summer season, due to the large overestimation of the precipitation shown in Figure 5a, whereas it is minimum (and negative) in the spring and in the fall, where all the other datasets have positive water balances.

Finally, we examine the seasonal cycle of the runoff. The runoff is the output of the redistribution of water within soil due to the water balance $B$, so that, as mentioned before, the runoff results to be a smoother function of time, with peaks and dips delayed with respect to those of $B$. In Figure 8a we show the monthly long-term accumulated runoff for all RCMs and reanalyses. Moreover, the seasonal cycle of the reconstructed actual runoff of Danube, computed following the strategy depicted in subsection 2d, and of the actual Danube discharge are depicted. All RCMs are in broad qualitative agreement with the reconstructed runoff, and feature a spring maximum and a late summer minimum. Note that, in all cases, the delay between the minimum of the water balance and the corresponding minimum of the runoff (~ 2 months) is shorter that the delay between the two maxima (~ 5 months). Such a nonlinear effect is due to the accumulation of water in the solid phase as seasonal snow cover during the winter months and to its subsequent rapid thawing in spring. The disagreement between RCMs is largest in this period of the year, suggesting that the representation of the snowpack is somewhat delicate. In Figure 8b, we synthesize the information contained in Figure 8a by depicting the absolute and relative (with respect to the mean value) amplitude of the seasonal cycle for all RCMs. The RCMs do not feature a high degree of quantitative agreement, spanning a relatively wide range of values for both measures of the amplitude of the seasonal cycle, and the clustering observed regarding the long-term annual statistics in Figs. 2, 3, and 4 is not reproduced here. When focusing on the verification side of the auditing processes, we observe that the two versions SMHI25 and SMHI50 are in remarkable agreement with the reconstructed runoff, whereas most RCMs simulate seasonal cycles with smaller amplitudes, both in relative and absolute terms. In Figs. 8a and 8b we also depict the information relative to the seasonal cycle of the observed discharge of the Danube river. As foreseen in subsection 2d, the actual discharge of a river results from an effective low-pass filtering of the integrated runoff, so that its seasonal cycle is a rather a slow-varying function - the amplitude of the seasonal cycle is more than halved - , where the peak and the dip are delayed, occurring in late spring and early fall, respectively. As already noted when commenting Figure 4, the two reanalyses do not provide any useful information regarding the statistical properties of the runoff in the Danube basin: in particular we may note that the NCEP/NCAR dataset provides a seasonal cycle having a heavily exaggerated amplitude, featuring also an unphysical summer relative maximum due to the summer maximum of water balance shown in Figure 7a.

## 5. Summary and Conclusions

In this paper we have intercompared and verified several RCMs in their representation of the statistical properties of the hydrological balance over the Danube river basin for the time frame 1961-1990. All of the analyzed RCMs have participated to the EU project PRUDENCE (http://dmi.prudence.dk) and are forced at the boundaries by the same run of the same driving AGCM – HadAM3 (scenario A2). For matters of completeness, we have also considered the outputs of HadAM3, and of the ECMWF and NCEP/NCAR reanalyses. The Danube has been chosen as a case study because of its multiple relevance at socio-economical as well as environmental and climatic level. While being well within continental Europe, the Danube basin has a direct relevance to the Mediterranean region, since it provides a relevant input of freshwater to the sea as well as being fuelled mostly by precipitations due to water of Mediterranean origin. Moreover, the basin is large enough (about 800.000 $Km^2$) to be well-resolved by RCMs and small enough with respect to their typical domains not to have statistical properties that are a-priori constrained by the driving AGCM. It might be argued that the Danube basin is geographically close to the eastern boundary of the domain of some of the considered RCMs. Since the hydrological cycle, as widely discussed in this paper, is essentially meteorological, and not geographical in nature, actually this is not a critical issue, since, speaking in meteorological terms, *the weather comes from west*. The hydrological balance has been computed in two different, but *in principle* equivalent ways. The first approach, which has a more meteorological *nuance*, relies on integrating over the area of interest the precipitation and evaporation fields and taking the difference. The second approach, which is more typically hydrological, relies on integrating the total runoff field. The equivalence dependn on the conservation of water within the climate subdomains. The verification data, given the impossibility of having reliable space-integrated estimates of precipitation *and* evaporation or runoff fields, due to their complex statistical nature, have been chosen as the1961-1990 monthly discharge time series at the Danube Delta, as provided by the GRDC.

When considering annual averages of the hydrological balance computed starting from the precipitation and evaporation fields, even if they are driven by the same run of the same AGCM, the RCMs are largely not self-consistent with each other, with a relative span larger than 50% of their esnsemble mean. Qualitatively, RCMs are clustered in three groups featuring, respectively, a hydrological balance much smaller than, statistically consistent with, and much larger than what simulated by the driving AGCM. The discrepancies on the mean value of the hydrological balance, which depends on the large-scale climatological net influx of water vapor into the basin, implies that, within their domains, the RCMs differ in the representation of relevant synoptic features such as storm tracks. - slight latitudinal variations in the preferred path of the water vapour rich-perturbations – and/or in the efficiency of the large scale precipitation. Nevertheless, the time series of the annual balances of the RCMs feature a positive correlation (close to 1) with that of the AGCM, thus showing that a large scale control at least on the variability is present.

The analysis of the basin-integrated, yearly accumulated precipitation and evaporation fields allows us to detect the presence of a precipitation-evaporation positive feedback process. For each RCM, the two time series have a very high (close to 1) positive correlation (except for the PROMES UCM model), and at the same time the RCMs with high values of yearly accumulated precipitation feature in most cases strong evaporation. In particular, in the SMHI and DMI models, increases in the resolution do not alter the net water balance, while speeding up the hydrological cycle by enhancing precipitation *and* evaporation.

The analysis of the yearly accumulated runoff allows for assessing inconsistencies within each RCM in the treatment of water within soil. All RCMs provide data which are consistent with the hypothesis that they conserve water when soil modeling is concerned, whereas one model (CLM GKSS) seems to *create* water within the soil, since the basin integrated long-term average of runoff is larger than that of water balance in a statistically significant way.

Large discrepancies among RCMs exist when the climatologies of the monthly accumulated fields are considered. For precipitation, the statistics of the model for the annual maxima is bimodal for both maxima and minima: most models have weakest precipitation in winter and strongest precipitation in early summer, whereas for some models the annual extremes are realized in late summer (dry) and fall (wet), respectively. Due to the dominant role played by the solar insolation, all RCMs agree on setting the maxima and minima of total evaporation nearby the summer and winter solstices, respectively. Nevertheless, also in this case, the quantitative span of models outputs is very large, ranging over 50%. When considering the annual balance, we have a good agreement among all RCMs, which all have a minimum of the balance in July and a maximum in November. The compensation between the anomalies in precipitation and evaporation due to the feedback as well as the large scale control due to the driving model may explain such an agreement.

In the case of the seasonal cycle of runoff, we have a surrogate of a verification data, since from the actual Danube discharge obtained from GRDC it has been possible to approximately regress statistically - following Hagemann et al. (2004) - the monthly basin-integrated runoff. Essentially all RCMs feature the same timing for maxima (late spring) and minima (fall), which is in broad agreement with the *quasi* observations, whereas quantitative disagreements exist, especially in the spring period, probably because of the problematic representation of the process of seasonal snowpack melting. When looking at the amplitude of the seasonal cycle, we observe that only two versions of the same model (SMHI25 and SMHI50) agree with quasi-observations, whereas most models tend to have a too flat seasonal cycle.

Finally, we find that the NCEP-NCAR and ERA-40 reanalyses are largely inadequate for representing the hydrology of the Danube river basin, both for the reconstruction of the long-term averages and of the seasonal cycle, and cannot in any sense be used as verification. The ERA40 long-term water balance reanalysis is one order of magnitude smaller than observations, with several years featuring an unphysical negative balance - the Danube basin resulting to be an exporter of water - , and a huge amount of water is created in the soil model module. The NCEP/NCAR reanalysis is much better in the representation of the long-term average of the water balance, and its soil model is a little more consistent for water conservation (but far from being perfect). Nevertheless, when looking into monthly climatologies, the NCEP/NCAR dataset performances are much worse: the summer precipitations are greatly exaggerated, and, in spite of the evaporative feedback, the water balance is positive and maximizes in summer time, which is quite unreasonable. The runoff water cycle in also completely off-the-track with respect to the RCMs and the quasi-observations.

Since the considered approaches relies on the mass conservation principle and bypasses the details of the air-land interface modeling, we propose that the atmospheric components of RCMs still face difficulties in representing the current water balance even on a relatively large scale, such as that of the Danube basin. We may infer that RCMs performances might be even more problematic in the representation of the hydrological balance on smaller river basins.

Criticalities in the parameterization of the microphysics of non-convective precipitation and, in the representation within the limited domains of synoptic scale atmospheric circulation patterns, which may also be influenced by the details of the boundary conditions, are good candidates for these issues. Note that the DMI and METNO-HIRHAM models, which give rather different results for all the analyzed statistical properties, are actually the same model at dynamical level, but differ in the choice of the limited domain and in the parameterization of microphysical processes.

Since for some RCMs the hydrological balance estimates obtained with the runoff fields do not precisely agree with those obtained via precipitation minus evaporation, some deficiencies of the land model in the conservation of water are also apparent.

As a conclusion, we may note that since the driving AGCM is in excellent agreement with the verification data, some RCMs seem to degrade the information provided by the large scale flow, once the local, downscaled information they produce is upscaled to an intermediate range between the minimum resolvable scale and the domain size. This emphasizes the fact that the downscaling and upscaling procedures do not commute and are *both* problematic. We suggest that these results should be carefully considered in the perspective of auditing RCMs and assessing their ability to simulate future climate changes, which might be problematic even if the driving is performed by an excellent GCM.

Future studies will include a similar analysis performed on GCMs run under various scenarios, in particular those considered by PCMDI (http://www.pcmdi-llnl.gov)

**ACKNOWLEDGMENTS**

The authors wish to acknowledge the efforts of the modeling groups participating to PRUDENCE for the generous provision of open-access data for analysts, and in particular O.B. Christensen for facilitating the access to the datasets and Hadley Centre for giving permission to access the HadAM3 A2 . The provision of data by the Global Runoff Data Center (http://grdc.bafg.de/) for the monthly Danube discharge and by NOAA for El Niño-Southern Oscillation indexes (http://www.cpc.ncep.noaa.gov/data/indices/) is also acknowledged. The authors also wish to thank R. Bojariu for providing NAO time series for the HadAM3 model. This paper has been produced in the context of the EU INTERREG IIIB CADSES programme project HYDROCARE.

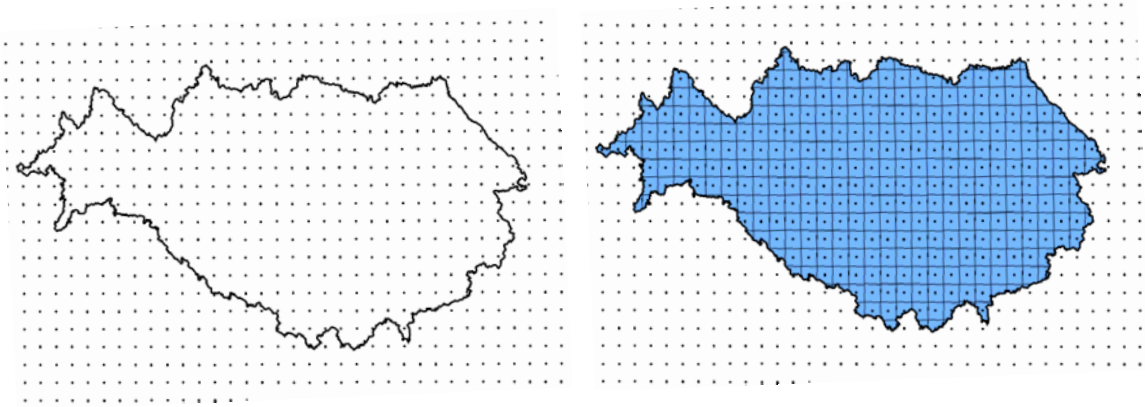

1a)                                            1b)

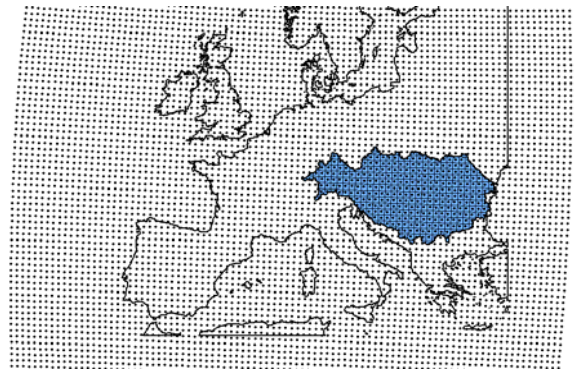

1c)

Figure 1. GIS data processing. 1a) point layer creation over the catchment area, 1b) point layer to Voronoi polygon layer transformation, 1c) the Lambert Azimuthal Equal Area projection of the catchment within Europe. The grid of the PROMES UCM model is used for explanatory purposes. Most models have higher resolution grids (see Table 1).

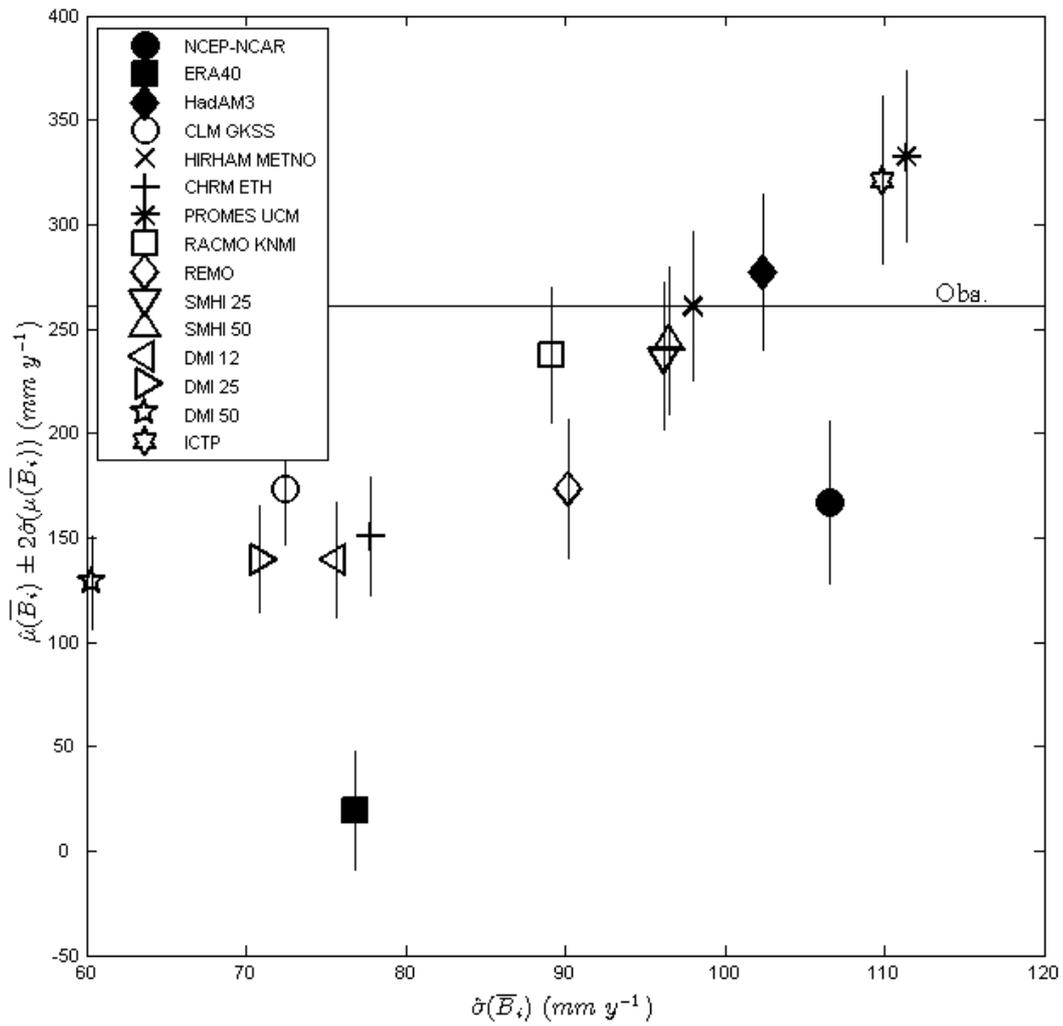

Figure 2. Basin-integrated annual accumulated water balance: interannual variability vs. 95% confidence interval of the mean. Obs. Stands for long-term averaged discharge at sea (about 6600 m$^3$s$^{-1}$). Note that 100 *mm y*$^{-1}$ of net water balance correspond to about 2500 *m$^3$s$^{-1}$* of equivalent mean river discharge. Further details in the text.

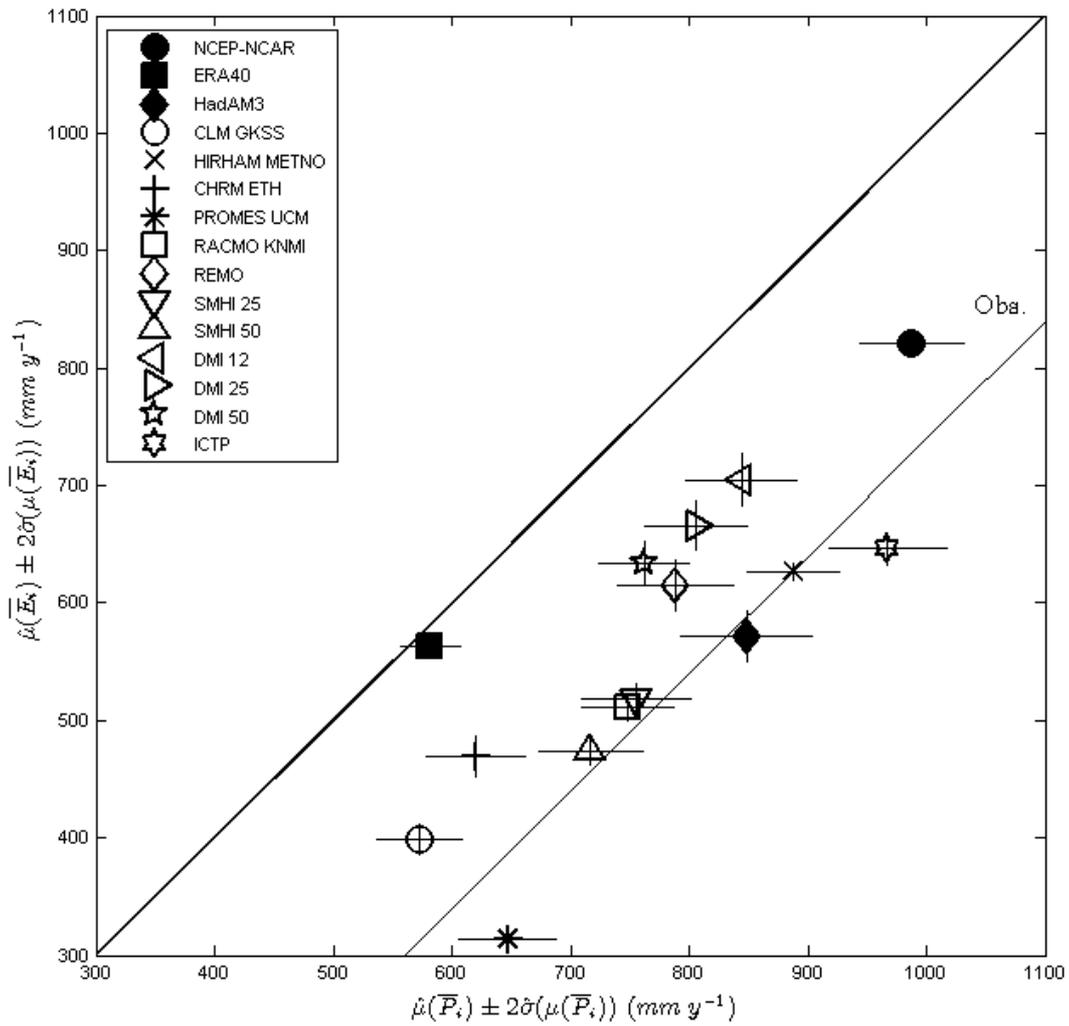

Figure 3. 95% confidence interval of the mean of the basin-integrated annual accumulated precipitation vs. the 95% confidence interval of the mean of the basin-integrated annual accumulated evaporation. The line captioned with Obs. describes the observative constraint given by the long-term averaged discharge at sea. The bisectrix gives the zero water balance case. Obs. Stands for long-term averaged discharge at sea (about 6600 $m^3 s^{-1}$). Note that 100 $mm\ y^{-1}$ of precipitation correspond to about 2500 $m^3 s^{-1}$ of equivalent mean river discharge. Further details in the text.

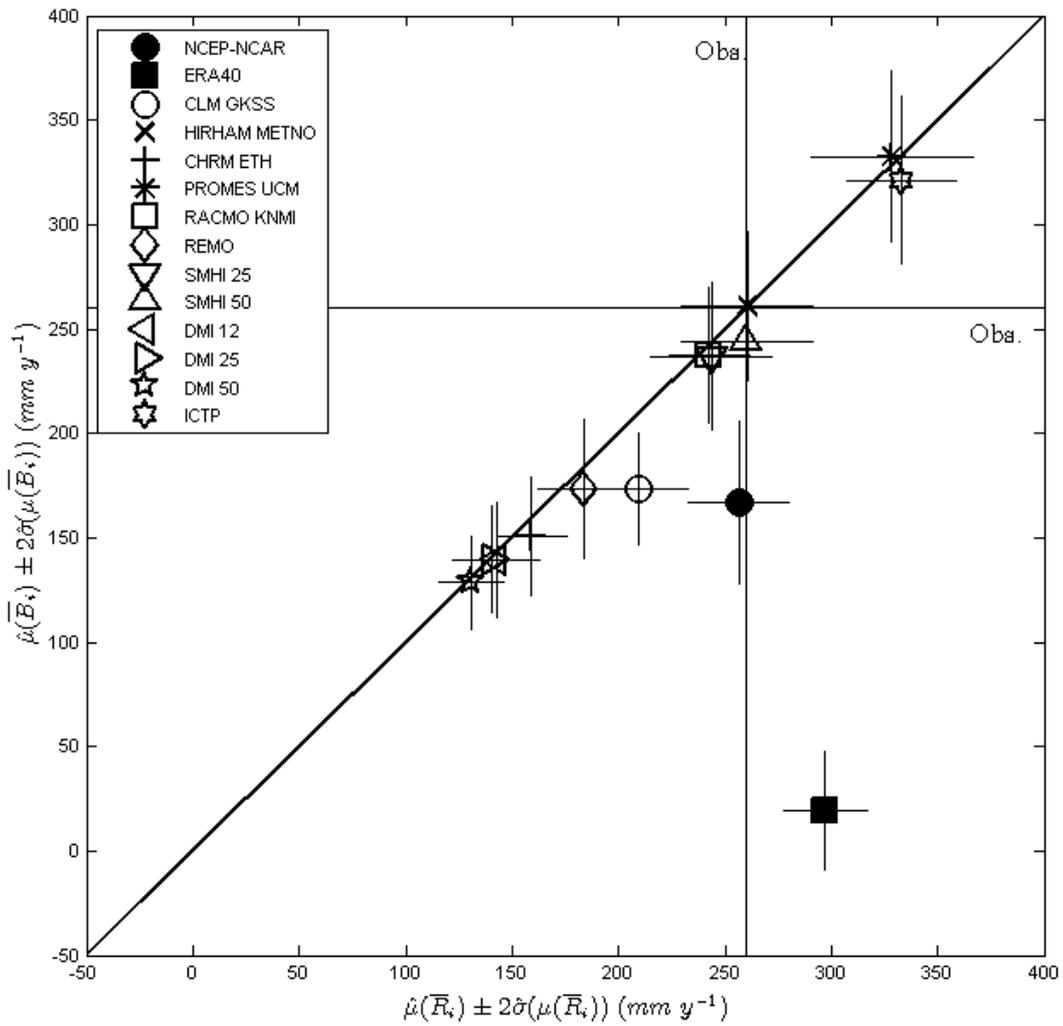

Figure 4. 95% confidence interval of the mean of the basin-integrated annual accumulated runoff vs. the 95% confidence interval of the mean of the basin-integrated annual accumulated precipitation net water balance. Obs. stands for long-term averaged discharge at sea (about 6600 $m^3s^{-1}$). The bisectrix indicates the theoretical constraint all datasets should obey to. Note that 100 $mm\ y^{-1}$ of net water balance (or of runoff) correspond to about 2500 $m^3s^{-1}$ of equivalent mean river discharge. Further details in the text.

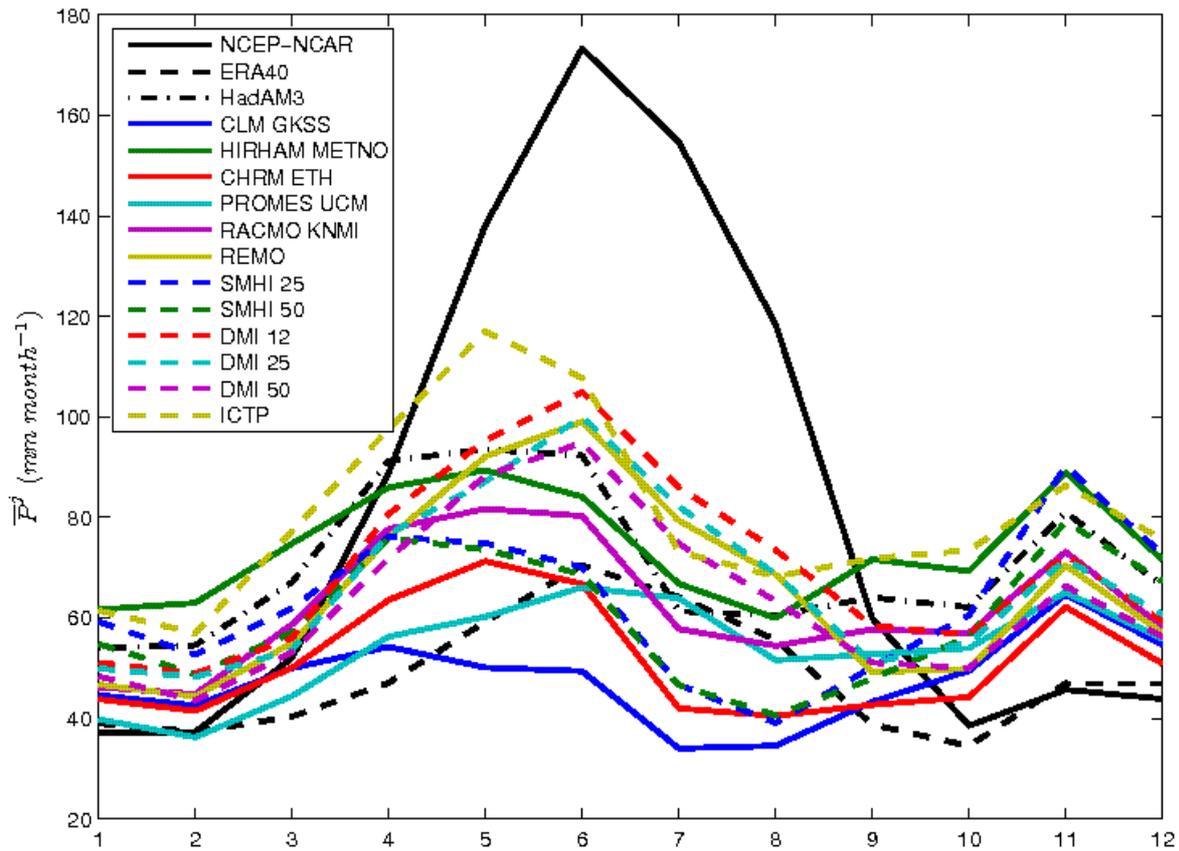

Figure 5a. Monthly long-term accumulated precipitation over the Danube catchment. Note that 10 *mm month*[-1] of precipitation correspond to about 3000 $m^3 s^{-1}$ of equivalent mean river discharge. Further details in the text.

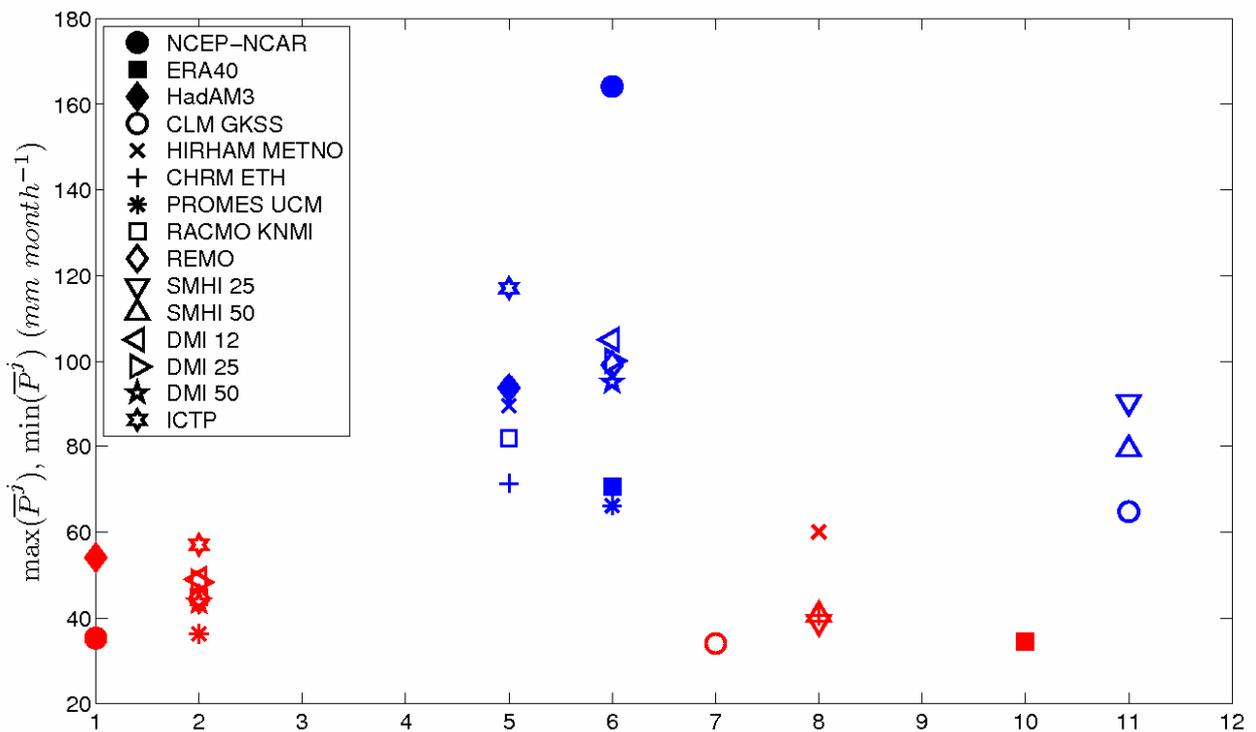

Figure 5b Occurrence of maxima (blue) and minima (red) of monthly long-term accumulated precipitation. Note that 10 *mm month*[-1] of precipitation correspond to ~ 3000 $m^3 s^{-1}$ of equivalent mean river discharge. Further details in the text.

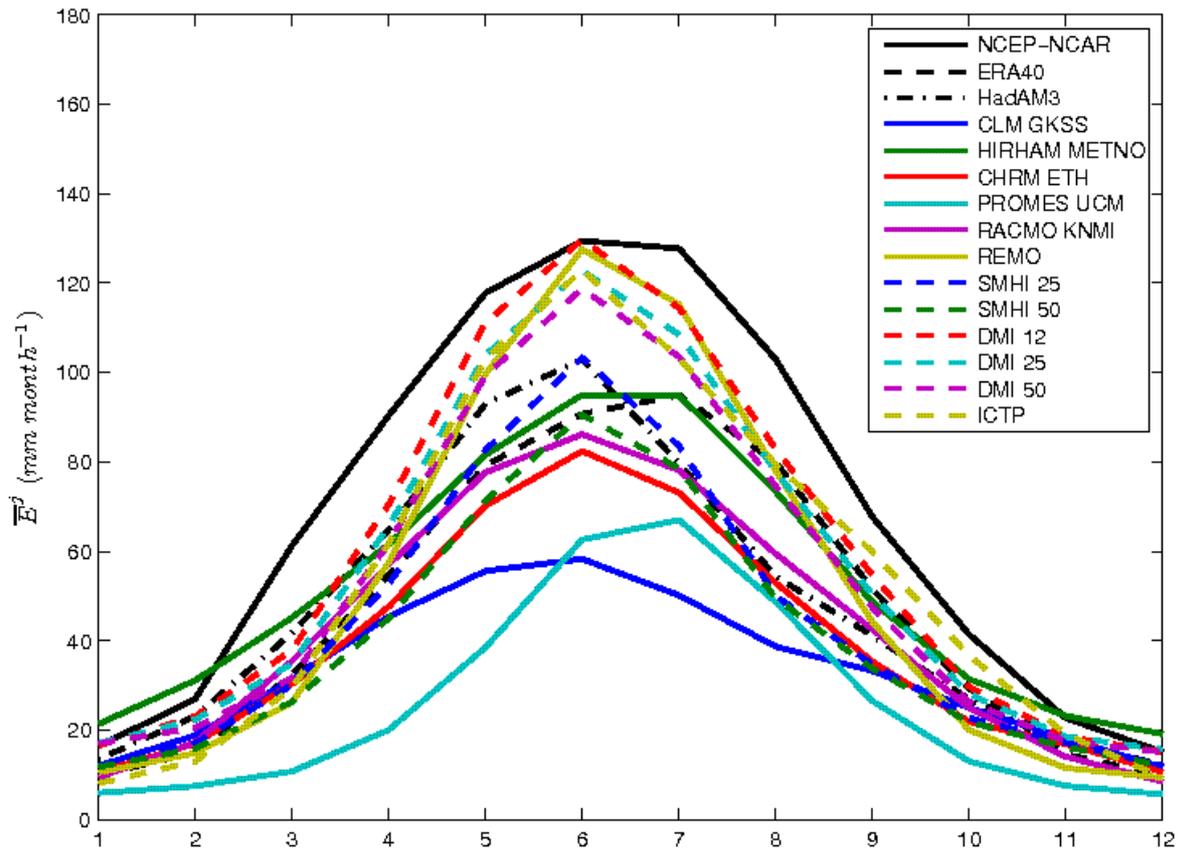

1
2  Figure 6a Monthly long-term accumulated evaporation over the Danube catchment. Note that 10 *mm month*$^{-1}$ of
3  evaporation correspond to about -3000 $m^3 s^{-1}$ of equivalent mean river discharge. Further details in the text.

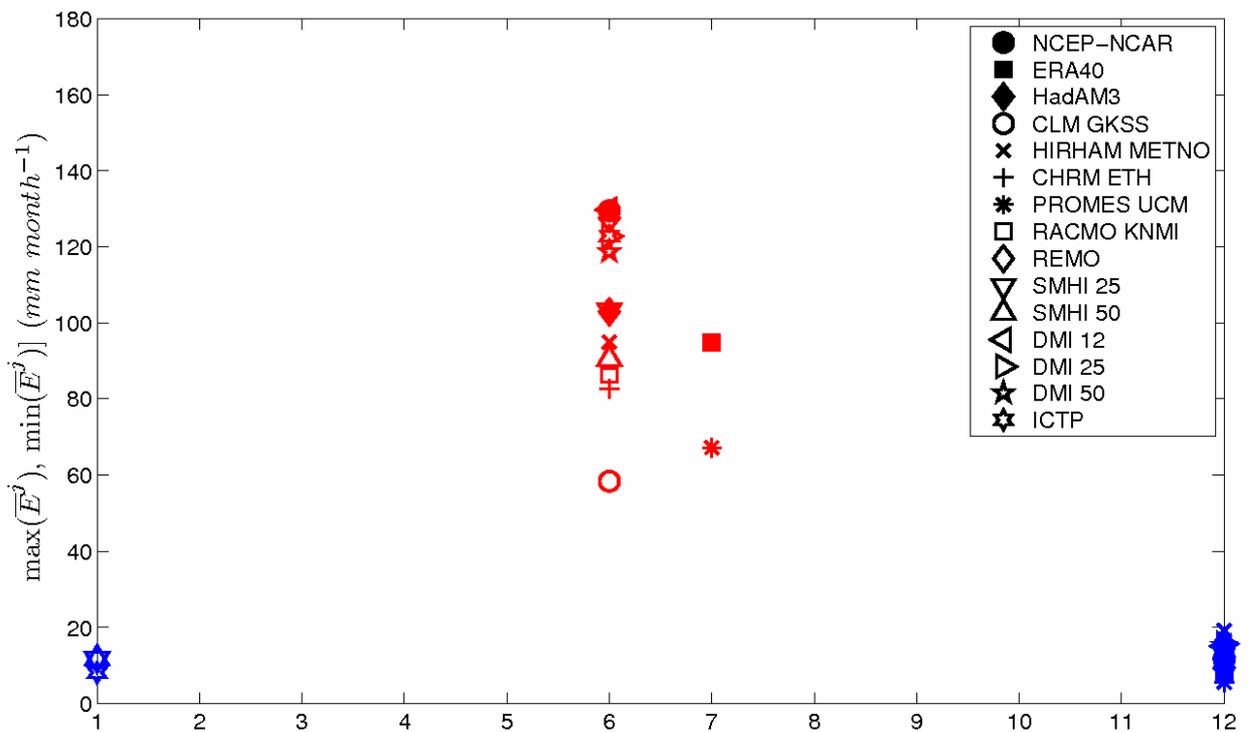

4
5  Figure 6b Occurrence of maxima (red) and minima (blue) of monthly long-term evaporation. Note that 10 *mm month*$^{-1}$
6  of evaporation correspond to ~ -3000 $m^3 s^{-1}$ of equivalent mean river discharge. Further details in the text.

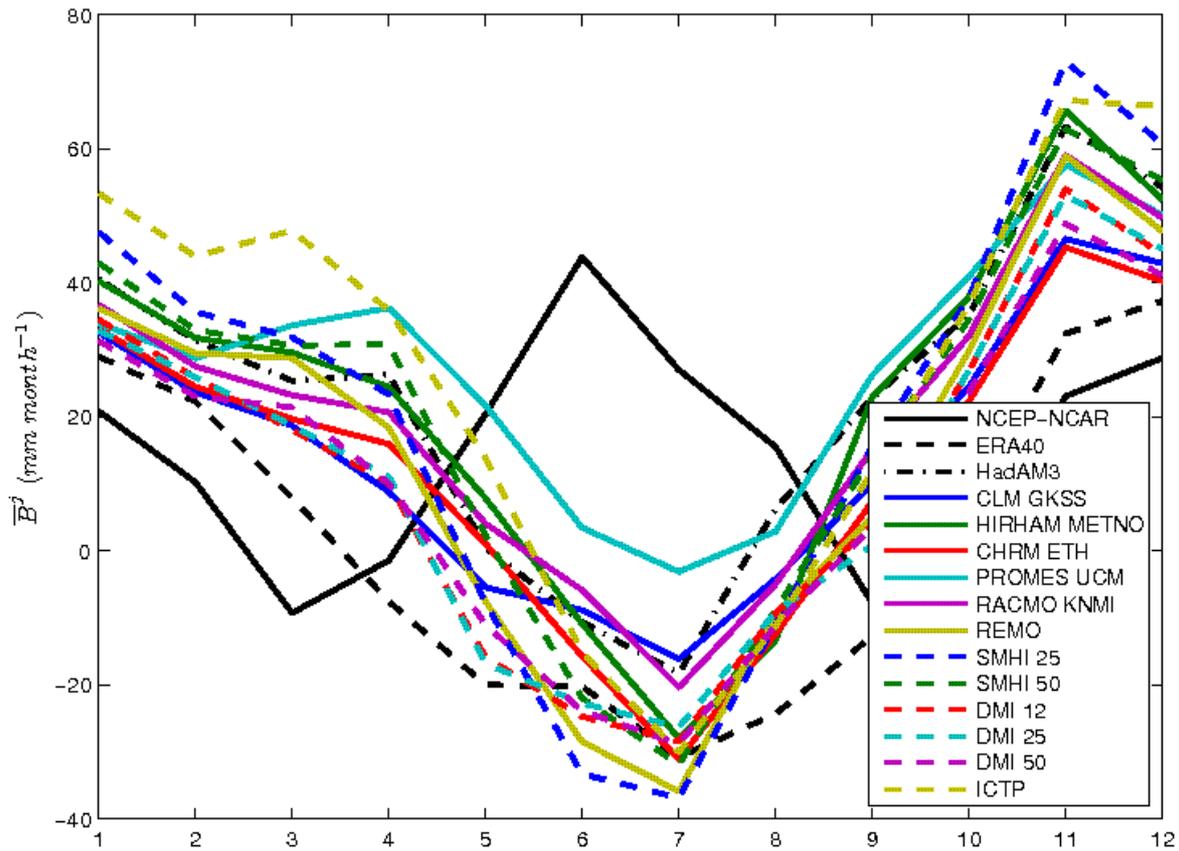

Figure 7a Monthly long-term accumulated net water balance over the Danube catchment Note that 10 *mm month*$^{-1}$ of net water balance correspond to about 3000 $m^3s^{-1}$ of equivalent mean river discharge. Further details in the text.

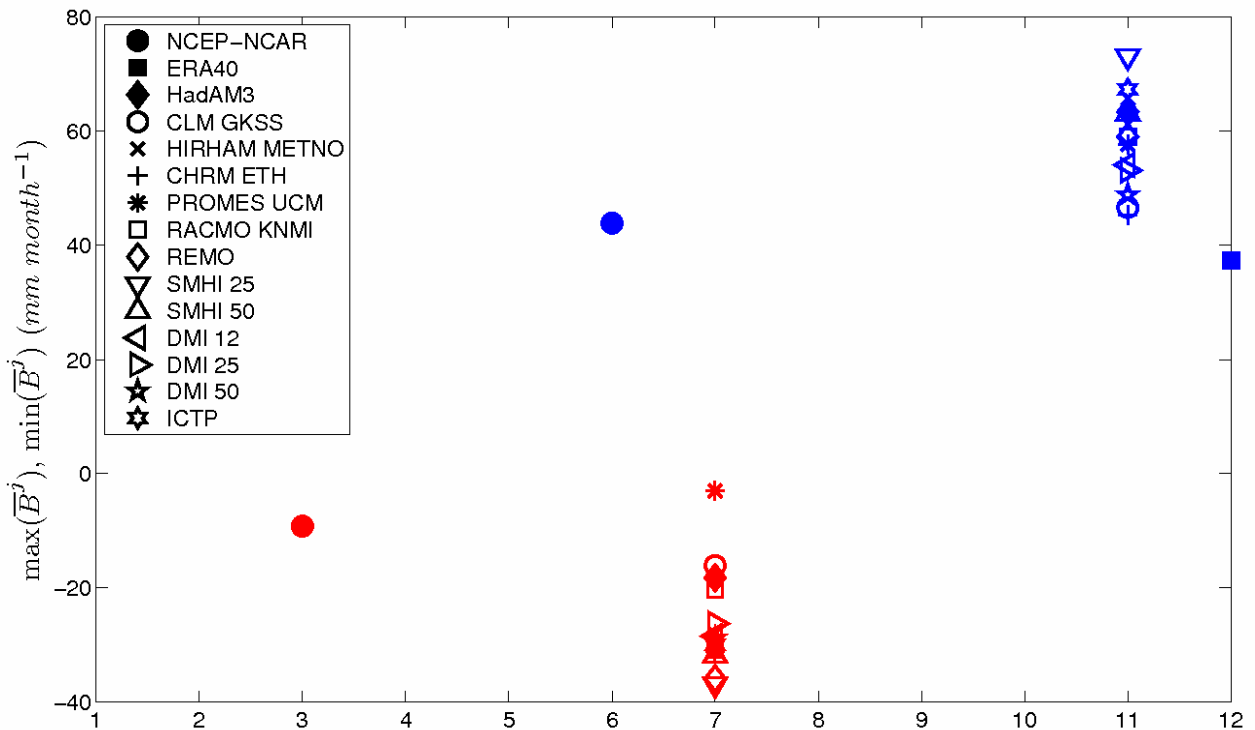

Figure 7b. Occurrence of maxima (blue) and minima (red) of monthly long-term net water balance. Note that 10 *mm month*$^{-1}$ of net water balance correspond to ~ 3000 $m^3s^{-1}$ of equivalent mean river discharge. Further details in the text.

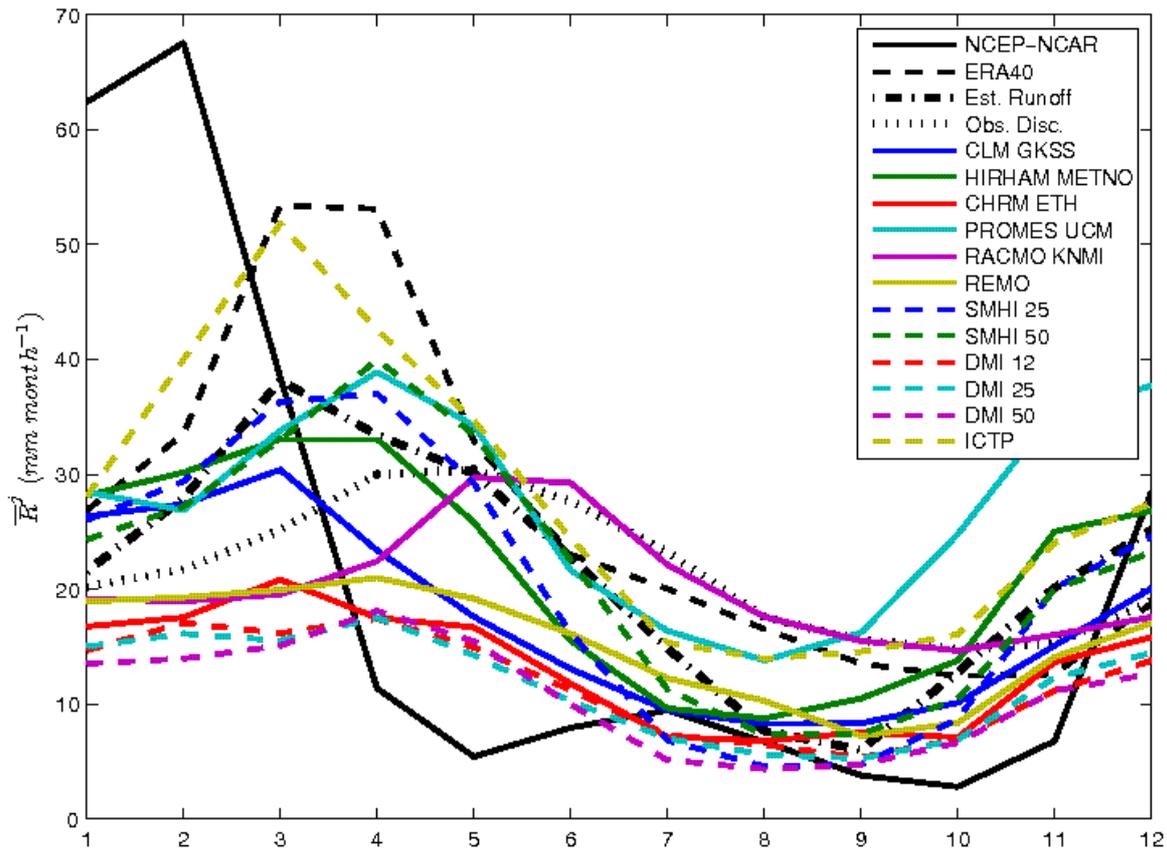

Figure 8a Monthly long-term accumulated runoff over the Danube catchment. Note that 10 *mm month*$^{-1}$ of runoff correspond to about 3000 *m$^3$s$^{-1}$* of equivalent mean river discharge. Further details in the text.

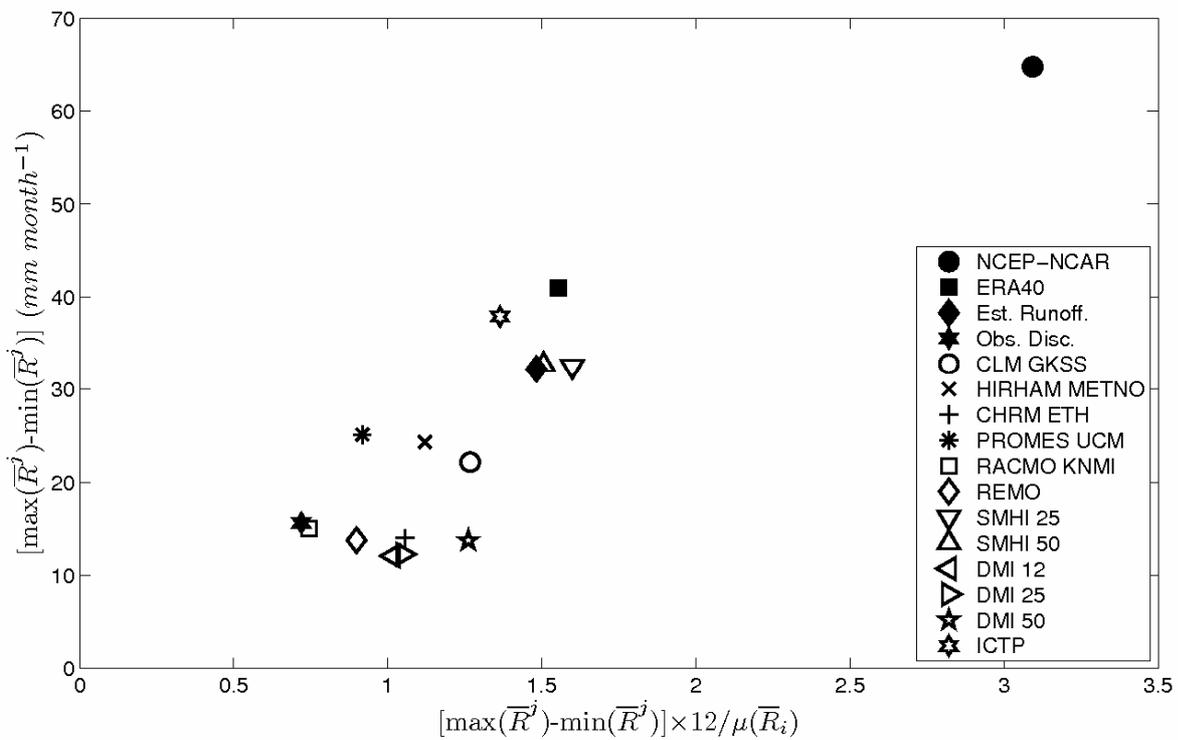

Figure 8b. Relative vs. Absolute amplitude of the seasonal cycle of the runoff over the Danube catchment. Note that 10 *mm month*$^{-1}$ of runoff correspond to about 3000 *m$^3$s$^{-1}$* of equivalent mean river discharge. Further details in the text..

Table 1: Overview of the Regional Climate Models considered in this study (1961-1990).

| Code | Model | Driving data | Institute | Country | Available data | lat x lon | Vertical levels |
|---|---|---|---|---|---|---|---|
| CLM GKSS | CLM | HadAM3H A2 | GKSS Research Centre Geesthacht | Germany | Daily | 0.50° x 0.50° | 20 |
| HIRHAM METNO | HIRHAM | HadAM3H A2 | Norwegian Meteorological Institute | Norway | Daily | 0.46° x 0.46° | 19 |
| CHRM ETH | CHRM | HadAM3H A2 | ETH - Swiss Federal Institute of Technology | Switzerland | Daily | 0.50° x 0.50° | 20 |
| PROMES UCM | PROMES | HadAM3H A2 | UCM - Universidad Complutense de Madrid | Spain | Daily | 0.50° x 0.50° | 26 |
| RACMO KNMI | RACMO | HadAM3H A2 | KNMI - The Royal Netherlands Meteorological Institute, University of Reading | Netherlands, UK | Daily | 0.44° x 0.44° | 31 |
| REMO | REMO | HadAM3H A2 | MPI - Max-Planck-Institute for Meteorology | Germany | Daily | 0.50° x 0.50° | 19 |
| SMHI25 | RCAO – high resolution | HadAM3H A2 | SMHI – Swedish Meteorological and Hydrological Institute | Sweden | Daily | 0.22° x 0.22° | 59 |
| SMHI50 | RCAO | HadAM3H A2 | SMHI – Swedish Meteorological and Hydrological Institute | Sweden | Daily | 0.44° x 0.44° | 24 |
| DMI12 | HIRHAM – extra high res. | HadAM3H A2 | DMI - Danish Meteorological Institute | Denmark | Monthly | 0.15° x 0.15° | 19 |
| DMI25 | HIRHAM – high resolution | HadAM3H A2 | DMI - Danish Meteorological Institute | Denmark | Daily | 0.22° x 0.22° | 19 |
| DMI50 | HIRHAM | HadAM3H A2 | DMI - Danish Meteorological Institute | Denmark | Daily | 0.44° x 0.44° | 19 |
| ICTP | ICTP –RegCM | HadAM3H A2 | ICTP The Abdus Salam Intl. Centre for Theoretical Physics | Italy | Daily | 0.44° x 0.44° | 23 |

Table 2: Overview of the other considered datasets (1961-1990).

| Code | Dataset | Institute | Country | Available data | lat x lon | Levels |
|---|---|---|---|---|---|---|
| ERA40 | ERA-40, T159 resolution – Reamalyses | ECMWF –European Center for Medium-Range Weather Forecast | UK | 4XDaily | 2.5° x 2.5° | 60 |
| NCEP-NCAR | NCEP-NCAR - Reanalyses | National Center for Environmental Prediction – National Center for Atmospheric Research | USA | 4XDaily | 1.905° x 1.875° | 28 |
| HadAM3 | HadAM3H model– A2 scenario (forced by observed SST and sea ice) | Hadley Centre for Climate Change - Met Office | UK | Daily | 1.25°x1.875° | 19 |
| Obs. Disc. | Danube discharge at Ceatal Izmail station | Global Runoff Data Center | Germany | Monthly | | |

Table 3. Values of model determined factor $a_i$ for 12 months (Hagemann et al. 2004)

| $i$ | 1 | 2 | 3 | 4 | 5 | 6 | 7 | 8 | 9 | 10 | 11 | 12 |
|---|---|---|---|---|---|---|---|---|---|---|---|---|
| $a_i$ | 0,83 | 1.04 | 0.93 | 0.81 | 0.94 | 0.95 | 1.05 | 1.22 | 2.09 | 2.54 | 1.23 | 0.97 |

Table 4: Correlation coefficients calculated for RCMs datasets vs. driving model datastes for yearly time series of integrated E, integrated P, and integrated water balance. In all cases the 95% confidence level for the null hypothesis of uncorrelation is about 0.4

| Dataset 1961-1990 – Yearly data | C(P,P) | C(E,E) | C(P-E,P-E) |
|---|---|---|---|
| HadAM3 | 1 | 1 | 1 |
| CLM GKSS | 0,91 | 0,66 | 0,92 |
| HIRHAM METNO | 0,90 | 0,53 | 0,92 |
| CHRM ETH | 0,87 | 0,70 | 0,89 |
| PROMES UCM | 0,87 | -0,16 | 0,90 |
| RACMO KNMI | 0,93 | 0,71 | 0,93 |
| REMO | 0,88 | 0,72 | 0,86 |
| SMHI 25 | 0,84 | 0,75 | 0,85 |
| SMHI 50 | 0,89 | 0,79 | 0,89 |
| DMI 12 | 0,85 | 0,73 | 0,86 |
| DMI 25 | 0,87 | 0,76 | 0,85 |
| DMI 50 | 0,80 | 0,75 | 0,78 |
| ICTP | 0,84 | 0,66 | 0,85 |

Table 5: Correlation coefficients of the yearly time series of the integrated values of E and P calculated for all datasets of models. In all cases the 95% confidence level for the null hypothesis of uncorrelation is about 0.4.

| Dataset 1961-1990 – Yearly data | C(E,P) |
|---|---|
| NCEP-NCAR | 0,73 |
| ERA40 | -0,41 |
| HadAM3 | 0,90 |
| CLM GKSS | 0,81 |
| HIRHAM METNO | 0,58 |
| CHRM ETH | 0,84 |
| PROMES UCM | 0,04 |
| RACMO KNMI | 0,69 |
| REMO | 0,82 |
| SMHI 25 | 0,87 |
| SMHI 50 | 0,89 |
| DMI 12 | 0,91 |
| DMI 25 | 0,90 |
| DMI 50 | 0,92 |
| ICTP | 0,80 |